\documentclass[conference]{IEEEtran}

\usepackage{array}
\usepackage[utf8]{inputenc}
\usepackage{amsmath, commath}
\usepackage{amsthm}
\usepackage{amsfonts}
\usepackage{amssymb}
\usepackage{bm}
\usepackage{bbm}
\usepackage{graphicx}
\usepackage{graphics}
\usepackage{xcolor}
\usepackage{float}
\usepackage{tikz}
\usetikzlibrary{shapes, snakes, patterns}

\usepackage{epstopdf}
\usepackage[export]{adjustbox}
\usepackage[list=true]{subcaption}
\usepackage{color}
\usepackage{algorithm}
\usepackage{algorithmic}

\usepackage[justification=centering]{caption}
\usepackage{enumerate} 
\usepackage{cite}
\usepackage{suffix}
\usepackage{mathtools}
\usepackage{tabularx, booktabs}
\usepackage{multirow}
\usepackage{multicol}
\usepackage{diagbox}
\usepackage{amsfonts}
\usepackage{amssymb}
\usepackage{textcomp}
\usepackage{dsfont} 
\usepackage{svg}
\usepackage{pgfplots,pgfplotstable}
\pgfplotsset{compat=1.3}
\usetikzlibrary{automata, positioning, arrows,shapes.multipart,fit}

\def\0{{\mathbf 0}}

\def\i1{{\mathbf 1}}

\usepackage{cite}
\usepackage{amsmath,amssymb,amsfonts}
\usepackage{algorithmic}
\usepackage{graphicx}
\usepackage{subcaption}
\usepackage{textcomp}
\usepackage{url}

\usepackage{array}
\usepackage{stfloats}

\usepackage{multirow} 
\def\BibTeX{{\rm B\kern-.05em{\sc i\kern-.025em b}\kern-.08em
    T\kern-.1667em\lower.7ex\hbox{E}\kern-.125emX}}

\title{An Open-Source Experimentation Framework for the Edge Cloud Continuum}
\author{\IEEEauthorblockN{Georgios Koukis\IEEEauthorrefmark{3}\IEEEauthorrefmark{2},
Sotiris Skaperas\IEEEauthorrefmark{1}\IEEEauthorrefmark{2}, 
Ioanna Angeliki Kapetanidou\IEEEauthorrefmark{3}\IEEEauthorrefmark{2},
Vassilis Tsaoussidis\IEEEauthorrefmark{3}\IEEEauthorrefmark{2},
Lefteris Mamatas\IEEEauthorrefmark{1}\IEEEauthorrefmark{2}}
\IEEEauthorblockA{\IEEEauthorrefmark{2} Athena Research and Innovation Center, Greece}
\IEEEauthorblockA{\IEEEauthorrefmark{3} Department of Electrical and Computer Engineering, Democritus University of Thrace, Xanthi, Greece}
\IEEEauthorblockA{\IEEEauthorrefmark{1} Department of Applied Informatics, University of Macedonia, Thessaloniki, Greece} 

Emails: \{gkoukis, ikapetan, vtsaousi\}@ee.duth.gr, \{sotskap, emamatas\}@uom.edu.gr 
}

\begin{document}
\maketitle
\begin{abstract}
The CODECO Experimentation Framework is an open-source solution designed for the rapid experimentation of Kubernetes-based edge cloud deployments. It adopts a microservice-based architecture and introduces innovative abstractions for (i) the holistic deployment of Kubernetes clusters and associated applications, starting from the VM allocation level; (ii) declarative cross-layer experiment configuration; and (iii) automation features covering the entire experimental process, from the configuration up to the results visualization. We present proof-of-concept results that demonstrate the above capabilities in three distinct contexts: (i) a comparative evaluation of various network fabrics across different edge-oriented Kubernetes distributions; (ii) the automated deployment of EdgeNet, which is a complex edge cloud orchestration system; and (iii) an assessment of anomaly detection (AD) workflows tailored for edge environments.   
\end{abstract}

\maketitle

\section{Introduction}

Edge computing has become a paramount paradigm to meet the requirements of the ever-increasing Internet of Things (IoT) applications and services that often require real-time processing. By bringing computation and storage closer to the data sources and end-users, edge computing contributes to reduced processing and response times, network load, and overall improved delivery and system performance. 

Nevertheless, edge devices may be resource-constrained in terms of computational power, memory and energy support. Therefore, edge solutions should be designed in a resource-efficient manner to achieve optimal performance. In this context, microservice-based solutions have gained significant popularity as, unlike traditional monolithic applications, they allow for offloading parts of the service to edge devices\cite{fu2021adaptive}, leading to improved resource efficiency. In particular, microservice-based applications consist of several loosely-coupled microservices, each being responsible for a specific, single-purpose part of the application's functionality. These microservices can be deployed, scaled, and updated independently, offering distinct advantages such as composable software design, programming heterogeneity, and simplified debugging \cite{gan2019open, zhao}.

Kubernetes (K8s), the de facto solution for service orchestration, facilitates automated service deployment and configuration, while also providing advanced scheduling capabilities \cite{carrion2022kubernetes, han}. However, being originally designed for traditional cloud environments, K8s might not be well-suited to support today’s microservice-based, time-critical applications in the edge \cite{Bohm}. In this light, several lightweight Kubernetes derivatives (e.g., K0s\footnote{https://k0sproject.io/}, K3s\footnote{https://www.rancher.com/products/k3s}, MicroK8s\footnote{https://microk8s.io/}) have been developed specifically for resource-constrained or low-footprint edge devices, aiming to better address the unique requirements of the edge \cite{koziolek2023lightweight}. Additionally, various platforms, both K8s and non-K8s focused have been proposed for the edge, such as KubeEdge \cite{kubeedge}, KubeOne\footnote{https://www.kubermatic.com/products/kubermatic-kubeone/edge/}, and Open Horizon \cite{lfedge}, 
each one focusing on particular aspects, e.g., resource consumption, device computational power, etc. The optimal K8s distribution for each case depends on the deployment scenario and the associated key driving factors.

Besides the selected K8s distribution, there is a growing need for intelligent, flexible and holistic manipulation of edge compute and network resources. Motivated by this, the CODECO project \cite{codeco} adopts a cross-layer adaptation approach, aiming to ensure enhanced performance for microservice-based applications across the entire Edge-Cloud continuum. CODECO introduces novel components, including ACM for automated configuration, deployment and monitoring of edge cloud resources, MDM for data workflow observability, SWN for scheduling and re-scheduling of application workloads, PDLC taking intelligent decisions for edge cloud orchestration, and NetMA providing network-awareness to CODECO and handling secure connectivity across pods.

From this point of view, an edge cloud continuum experimentation framework should be able to incorporate holistic experimental definitions for all layers of the protocol stack and main K8s features, including the support of alternative edge infrastructures (e.g., choice of hypervisor and server hardware) and virtualization technologies, to accommodate the diversity of edge cloud demands \cite{valsamas2}. Other important requirements include: i) appropriately designed abstractions to reduce experimentation complexity, ii) advanced automation and reliability, enabling the execution of experiments with different combinations of cross-layer parameters, that may also serve as reference points to evaluate a diverse set of intelligent algorithms, spanning from machine learning (ML) to optimization techniques, and, iii) abstractions apt to support modular deployment extensions. 




Existing open testbed solutions, incorporate cutting-edge technologies, such as 5G/6G, distributed/cloud computing, ML/AI and optical technologies, that enable a wide range of experimentation activities \cite{PAPADOPOULOS201786}.  Foundational paradigms of such testbeds include the EU-located SLICES testbed \cite{slices-testbed} and the US-located FABRIC \cite{fabric}, both recognized for their large-scale capabilities. A similar approach is adopted by PlanetLab \cite{planetlab} where users and institutes can contribute nodes, focusing on the federation of the provided heterogeneous infrastructures. An evolution of Planetlab is EdgeNet \cite{edgenet2}, which targets globally distributed edge-cloud environments. EdgeNet utilizes native K8s for the deployment and node contribution processes, extending the latter with custom resources (CRs).

Another K8s-centered solution is our recently introduced ClusterSlice \cite{clusterslice} platform, which is a zero-touch solution for transforming testbed resources into fully operational K8s slices. The CODECO Experimentation Framework extends the novel Resource and Infrastructure Manager operators of ClusterSlice to be used as non-K8s components. These two abstractions handle the manipulation of VMs and compute resources, respectively. Compared to ClusterSlice, it is lightweight (i.e., requires only docker for execution) as well as it supports additional edge capabilities and environments (e.g., EdgeNet) and experimentation automation features (i.e., through its own innovative abstractions). 

CODECO Experimentation Framework is based on a microservice-based architecture that offers: 
\begin{enumerate}
\item holistic experiment configuration, including the deployed infrastructure and K8s configuration parameters as well as the input of the experiments to be conducted,
\item innovative, modular and extensive abstractions, to accommodate a diverse set of experiments based on Ansible playbook templates, 
\item complete experimentation automation, from cluster deployment to experiment execution and results processing.
\end{enumerate}

Along these lines, the framework possesses additional advantages: i) reproducible experimentation with one-liner commands and versioning of components, ii) minimal operational support and risk since the deployments can be easily deleted and then re-configured, iii) integration with external edge environments and testbeds, e.g., EdgeNet, Cloudlab etc. The novel CODECO Experimentation Framework's capabilities are demonstrated through three proof-of-concept experiments, ranging from evaluating different network plugins across various K8s distributions to deploying EdgeNet software and assessing various anomaly detection (AD) approaches.

The remainder of the paper is structured as follows. In Section \ref{design}, we detail the architecture of CODECO Experimentation Framework along with its individual components. In Section \ref{results}, we give our proof-of-concept results, highlighting the main features and capabilities of the framework. Finally, Section \ref{concl} outlines our conclusions and future plans.

\section{Design and Implementation}\label{design}
Here, we present the CODECO Experimentation Framework. Specifically, we describe a simple experiment definition file and give an overview of the framework's architecture.

\subsection{Experiment Descriptor}\label{exper-descr}

The framework encompasses the entire experimentation process, beginning with the declarative definition of experiments for cluster development and the intended experiments. It then proceeds to the automated deployment of clusters and applications, the execution of experiments based on defined metrics, and ultimately, to the results' acquisition and processing.

Fig. \ref{fig:example-definition} gives an example experiment descriptor. It includes parameters for the infrastructure configuration, considering cluster information such as cluster name, IP and hypervisor, OS image, IP and MAC addresses for each cluster node (master or worker). In addition, specific K8s configuration settings can be defined, such as K8s flavor and version, scheduler or networking plugin; applications to be deployed, such as monitoring tools, Docker, and multi-cluster solutions; or experiment parameters, including metrics, experiment controller to support (i.e., container identifying the particular experiment), number of replications, and output file creation. 

\begin{figure}
    \setlength{\columnsep}{0.2em} 
        \centering
        \fontsize{8pt}{10pt}
        \begin{verbatim}
# generic configuration 
experiment_title="my_experiment"
username=athena
password=<ENCODED_PASSWORD>     
# infrastructure configuration
infra_manager_ip=83.212.134.23
infra_manager_name=codecocloud
infra_manager_type=xcp-ng
use_snapshots=true
node_osimage=ubuntu-22-clean
master_hosts=["athm1"]
master_ips=["83.212.134.27"]
master_macs=["66:16:91:ec:09:03"]
worker_hosts=["athw1"]
worker_ips=["83.212.134.35"]
worker_macs=["66:16:91:ec:09:11"]
# kubernetes configuration 
k8s_type=vanilla  
k8s_scheduler=swm
k8s_networkfabric=l2s-m
k8s_version="1.23"
# application configuration
app_names=["docker", "dashboard"]
app_scopes=["all", "cluster"]
# experiment definition 
exp_manager=cni-plugins
exp_input=["k8s-flannel", "k3s-calico"]
exp_metrics=["latency"]
replications_number=10
results_output="PDF"
        \end{verbatim}
    \caption{A Simple Experiment Definition File}
    \label{fig:example-definition}
\end{figure}

\subsection{Architecture}
\begin{figure*}
  \centering
\includegraphics[width=4in]{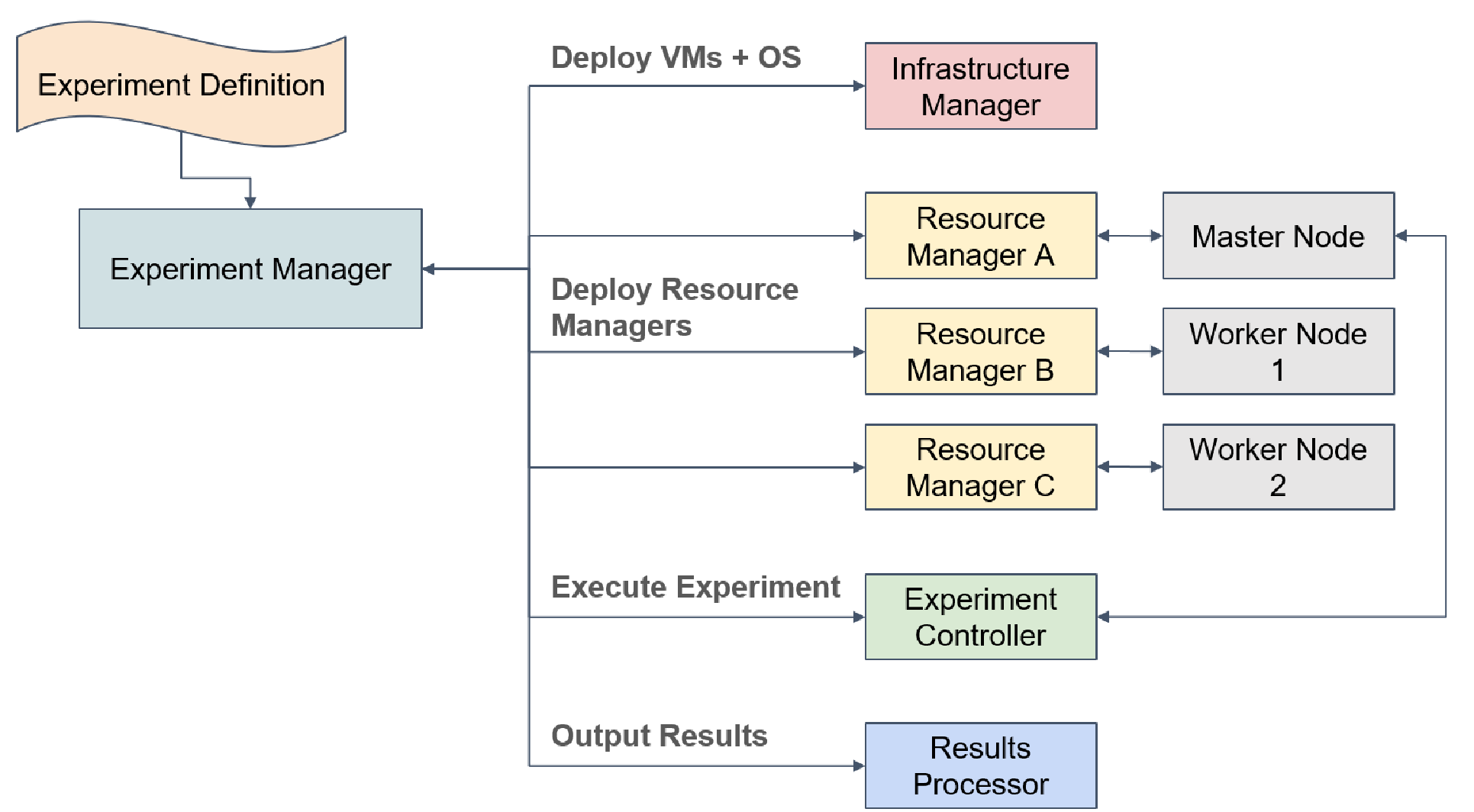}
  \caption {Main architectural components and interactions.}
  \label{fig:architecture}
\end{figure*}

The architecture of CODECO Experimentation Framework comprises the following components, as shown in Fig. \ref{fig:architecture}. All components take the form of microservices (i.e., containers) and can be independently extended to accommodate additional technologies or features. 


The \textit{Experiment Manager} oversees and coordinates all experiment processes, maintaining smooth execution without unexpected issues, such as ensuring reproducibility and facilitating automatic deployment and control of experiments.

The \textit{Infrastructure Manager} realizes a technology-agnostic abstraction over heterogeneous test-beds and cloud systems, which is responsible to allocate the cluster nodes from physical node or VM allocation to OS installation phases. In particular, specifications related to the allocation of nodes (e.g., number of master/worker nodes, OS image and snapshot usage) are the inputs of Infrastructure Manager, which in turn returns the addresses of allocated physical nodes or VMs. After the allocation of cluster nodes, a number of \textit{Resource Managers} are being deployed, one for each node.  

The \textit{Resource Managers} provide a node-level automation abstraction for the software deployment of cluster nodes. Specifically, they oversee the deployment of user-defined applications, supporting a range of them through the utilization of Ansible playbook templates (e.g., Docker, K8s dashboard, Liqo, Submariner, Argo Workflows, etc.), while also determining their versioning and deployment scope (e.g., deployment at the cluster level, across all servers etc.).

The component responsible for executing specific experiments once the cluster is operational is the \textit{Experiment Controller}. It offers a straightforward and automated experimentation approach, facilitating -among others- the performance evaluation of various CNI plugins and AD methods. The user is only required to input the experiment definition with the name of particular \textit{Experiment Controller} to use, specific metrics, along with the desired number of replications. Subsequently, the framework executes the experiments and produces the corresponding results.

Finally, the \textit{Results Processor} evaluates and post-processes the results of the experiments. Its tasks include performing statistical evaluations (e.g., mean values assessment), plotting the results based on the pre-defined metrics, and automating the creation of LaTeX-based report PDF files, incorporating the plots generated from the experiments.

\section{Proof-of-concept Results}\label{results}
This section details our experimentation setup and considered scenarios. In this regard, our proof-of-concept results demonstrate three distinct cases CODECO Experimentation Framework is utilized. In particular, we examine the: i) performance of network plugins across standard and Edge-oriented K8s distributions, ii) automatic deployment of a complicated edge system, i.e., EdgeNet, and iii) implementation of AD methods as on-demand K8s workflows.

\subsection{Experimentation Setup}\label{experiment-setup}
We execute the experiments in two testbeds referred to as ATH and UOM, which are located at the ATHENA Research Center and the University of Macedonia, respectively. Both testbeds use the XCP-ng virtualization platform\footnote{https://xcp-ng.org}. 
The former consists of a single Dell PowerEdge T640 physical machine equipped with an Intel(R) Xeon(R) Silver 4210R processor running at 2.40GHz, a 16-core CPU and 64GB RAM. The latter testbed comprises two Dell PowerEdge R630 physical servers, each one with 2 Intel(R) Xeon(R) CPU E5-2620 v4 processors operating at 2.10GHz with 16-core CPUs. 
\begin{figure*}[t]
\centering
   \begin{subfigure}[b]{0.32\textwidth}       \includegraphics[width=\textwidth]{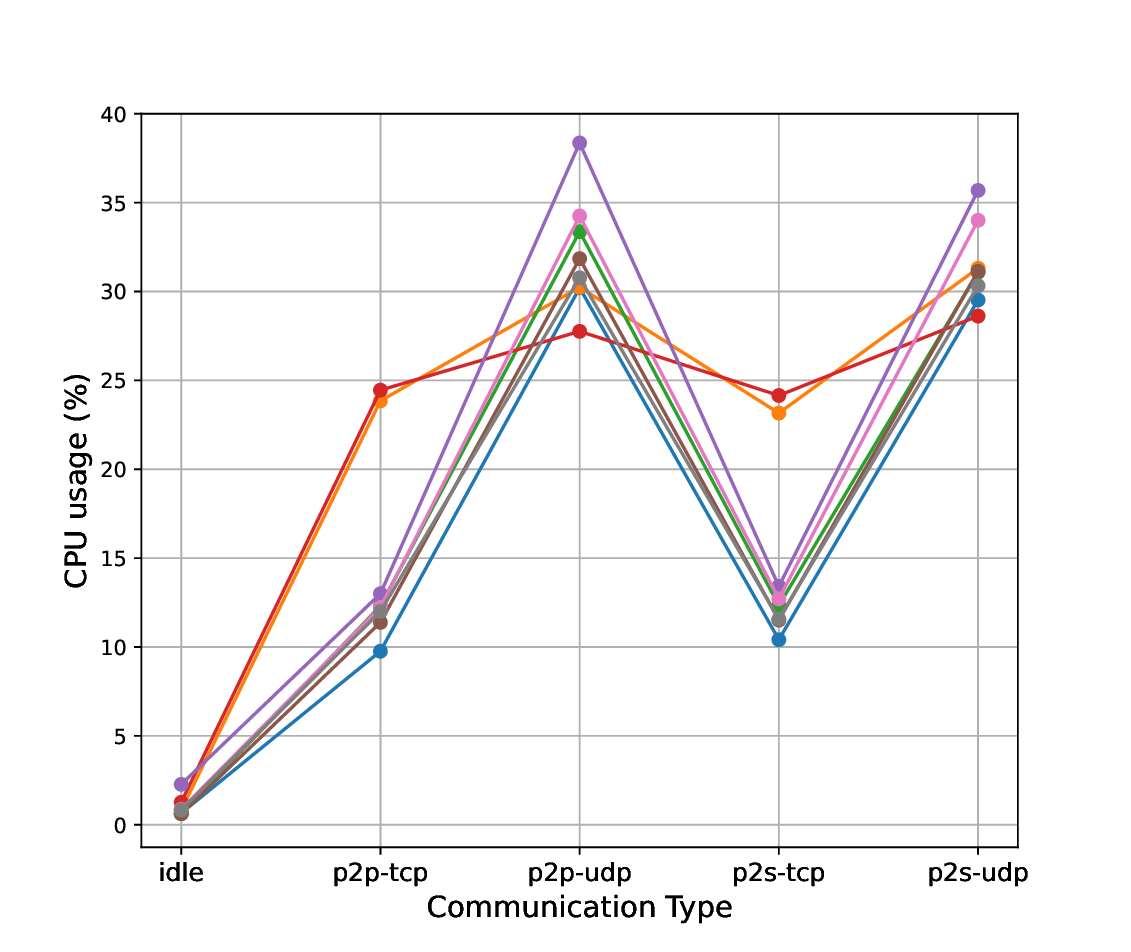}
   \caption{CPU utilization}
   \end{subfigure}
   \begin{subfigure}[b]{0.32\textwidth}       \includegraphics[width=\textwidth]{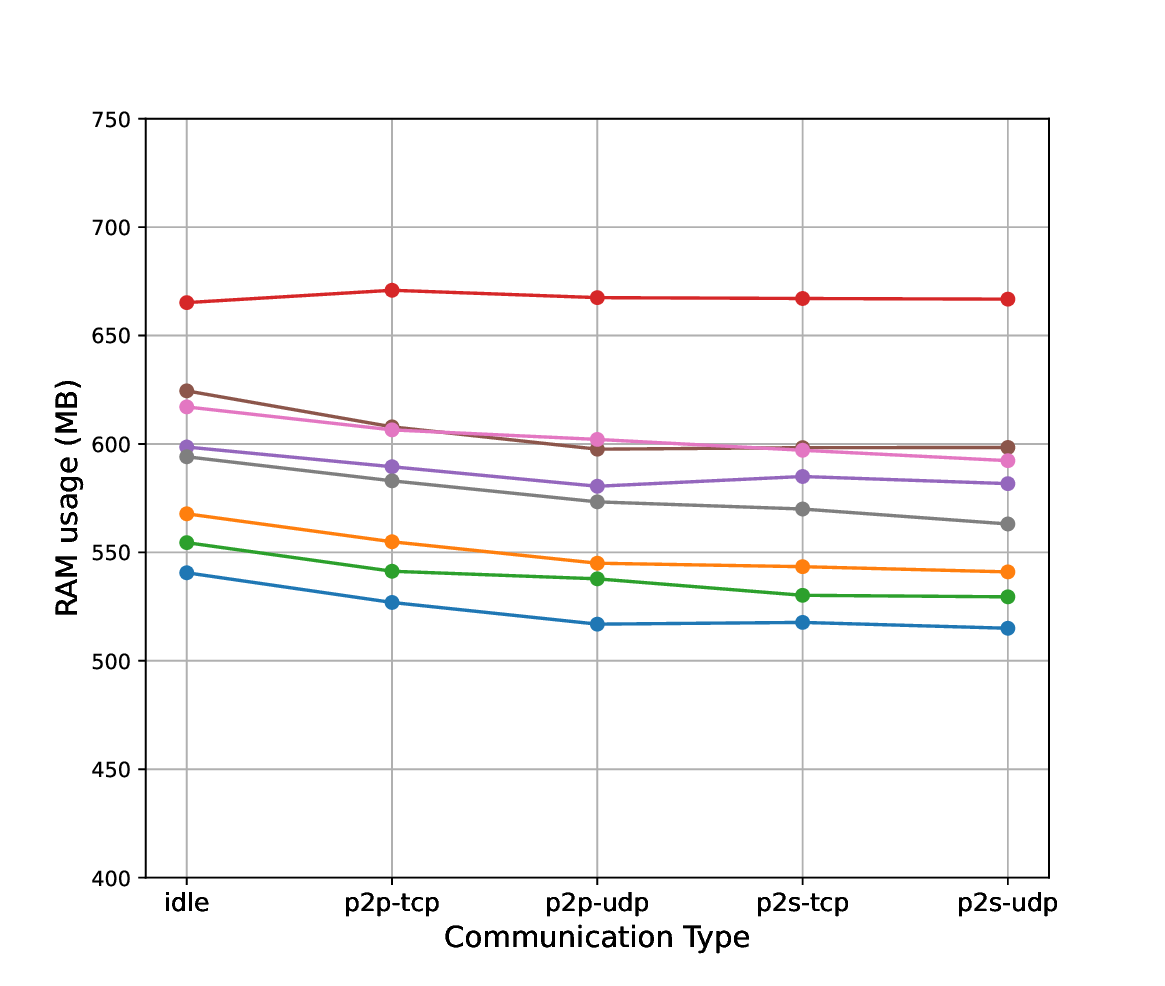}
   \caption{RAM consumption}
   \end{subfigure}%
   \begin{subfigure}[b]{0.34\textwidth}       \includegraphics[width=\textwidth]{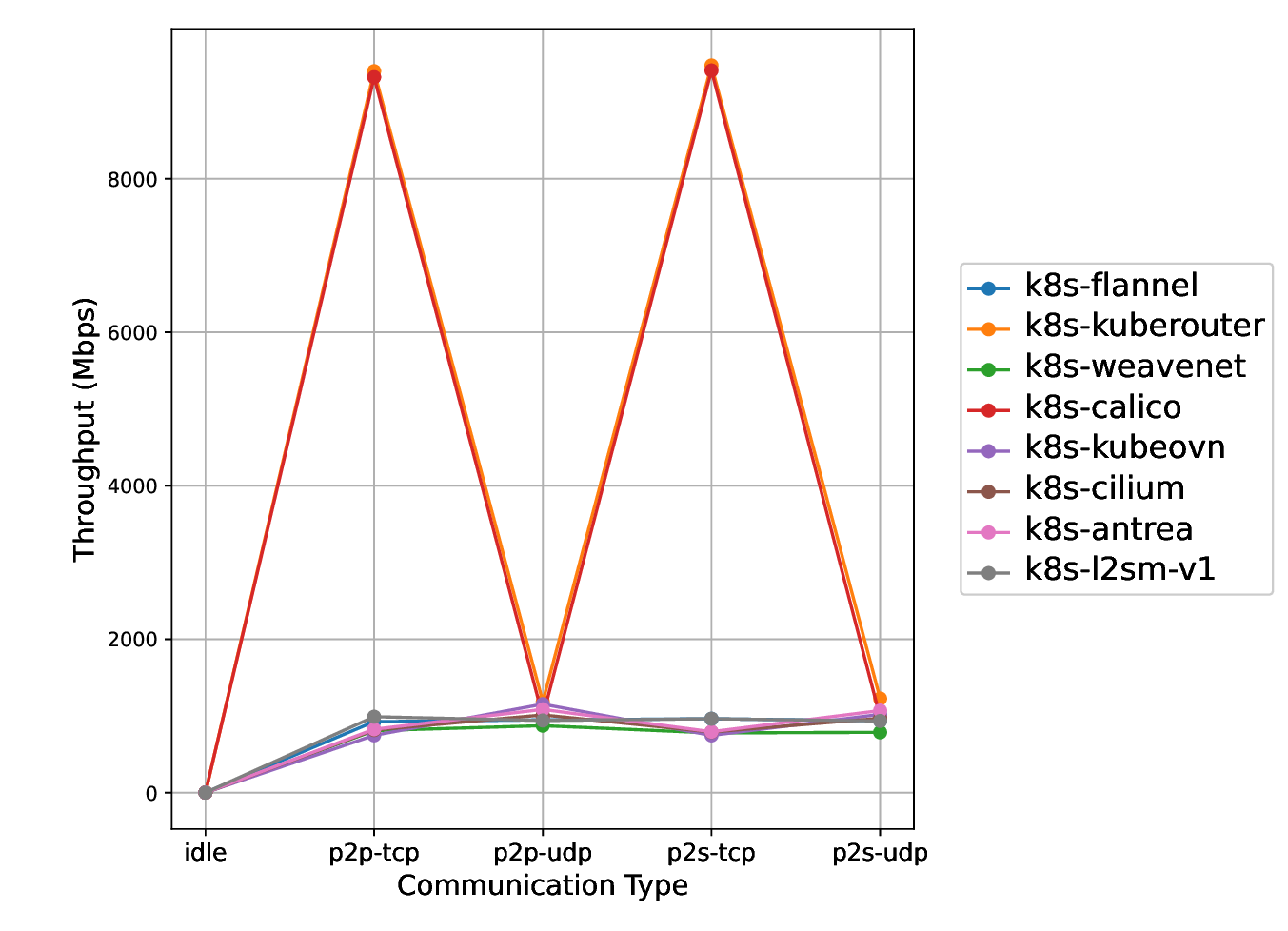}
   \caption{Achieved throughput}
   \end{subfigure}%
   \caption{CPU, RAM usage and throughput per CNI plugin, for the K8s distribution.}
   \label{plugins-metrics1}
\end{figure*}

\begin{figure*}[t]
\centering
   \begin{subfigure}[b]{0.32\textwidth}       \includegraphics[width=\textwidth]{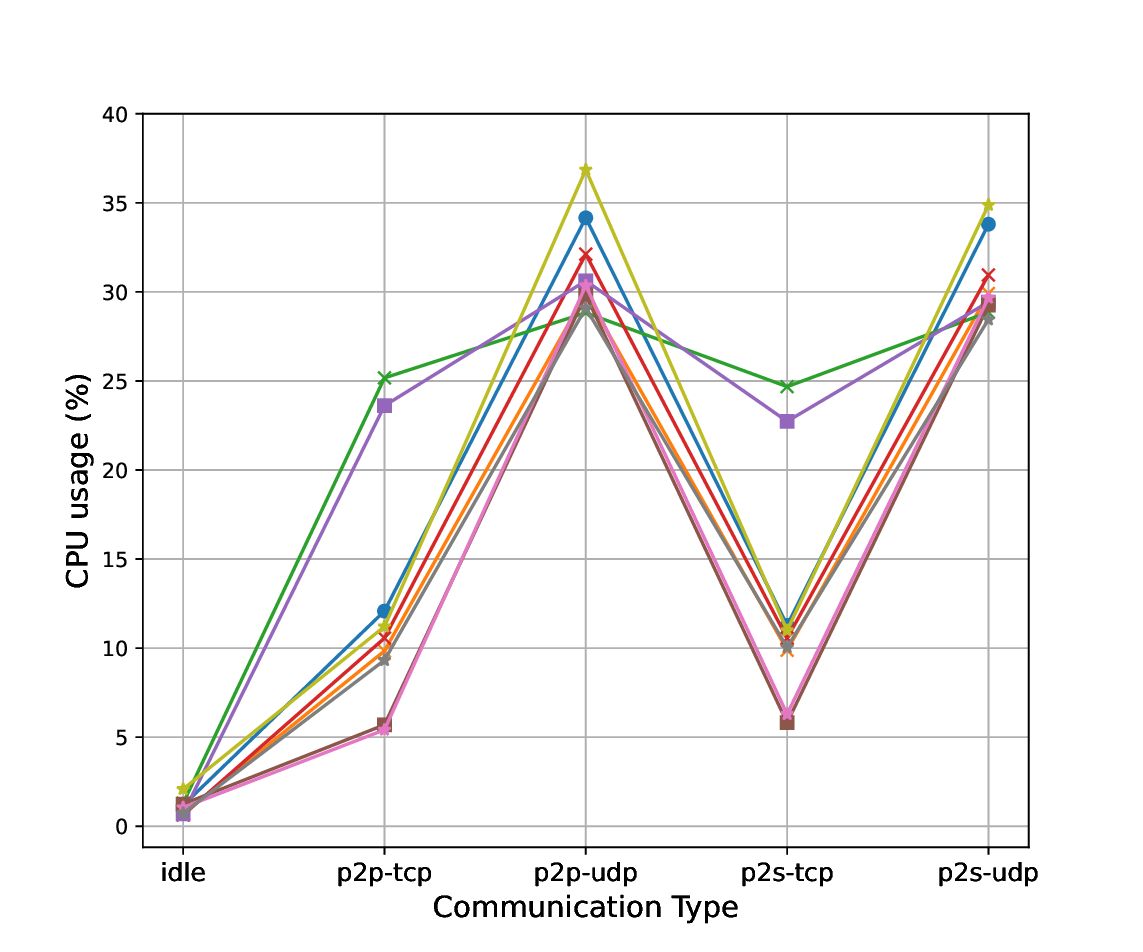}
   \caption{CPU utilization}
   \end{subfigure}
   \begin{subfigure}[b]{0.32\textwidth}       \includegraphics[width=\textwidth]{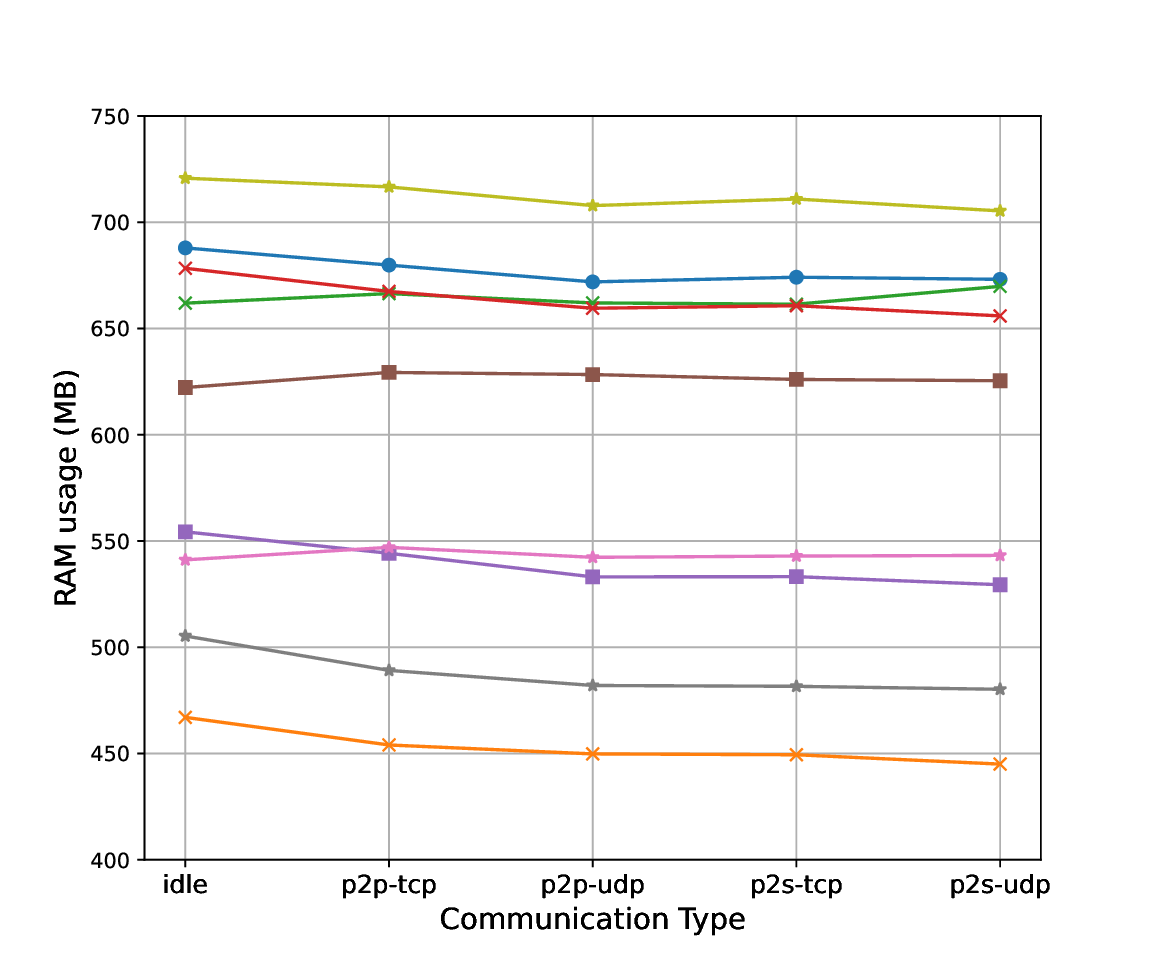}
   \caption{RAM consumption}
   \end{subfigure}%
   \begin{subfigure}[b]{0.345\textwidth}      \includegraphics[width=\textwidth]{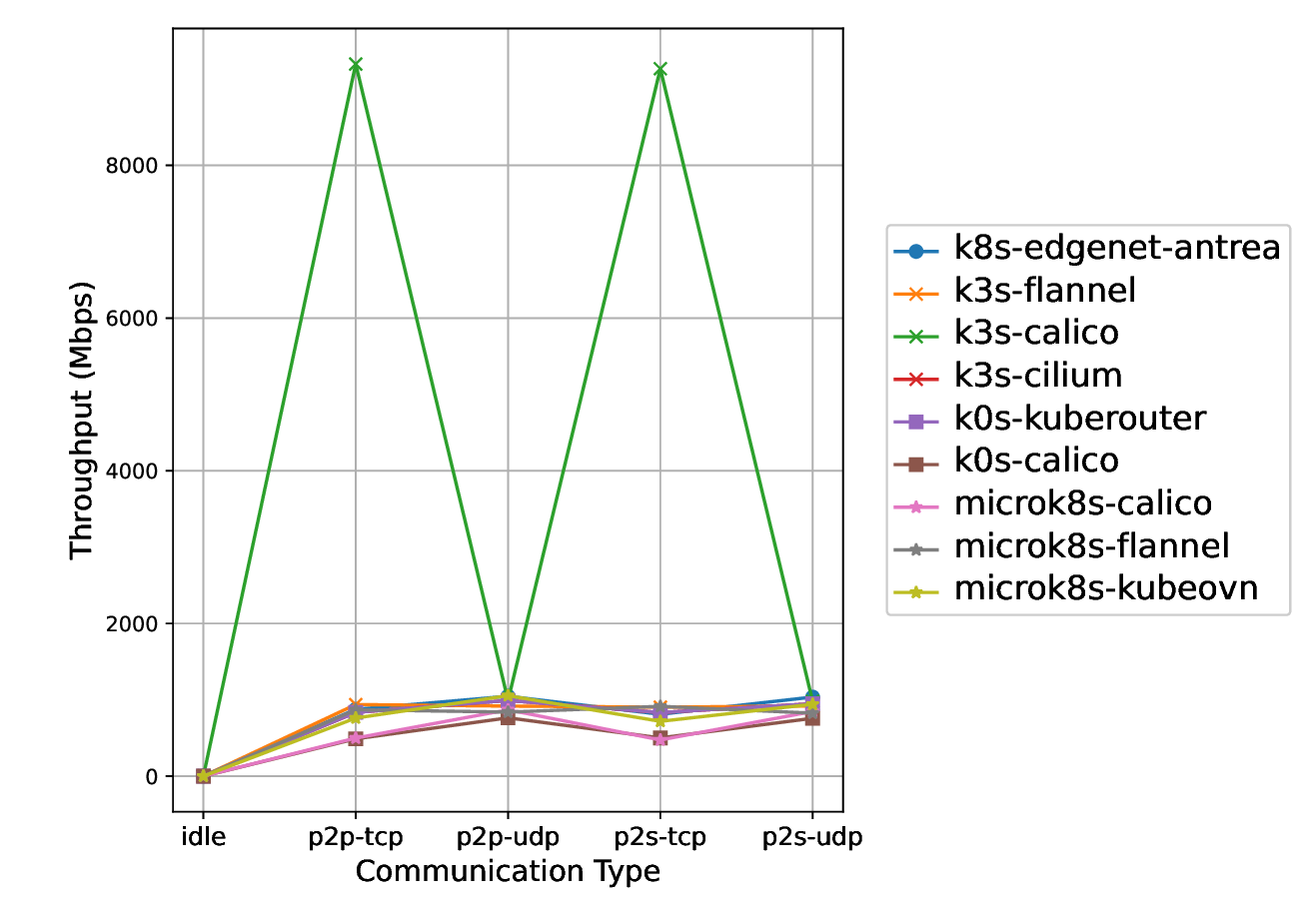}
   \caption{Achieved throughput}
   \end{subfigure}%
   \caption{CPU, RAM usage and throughput per CNI plugin, for lightweight distributions.}
   \label{plugins-metrics2}
\end{figure*}

\subsection{Experimentation Scenarios}
\subsubsection{Edge-oriented K8s Distributions and Network Plugins}\label{plugins&distros}
The CODECO Experimentation Framework supports a variety of K8s distributions and network plugins ranging from resource-intensive, production-based and feature-rich solutions to more lightweight and resource-constrained approaches.

In particular, the framework supports four widely used K8s distributions, i.e., vanilla K8s, K3s, K0s and MicroK8s. The feature-rich vanilla K8s is the standard K8s flavor, while the rest are lightweight distributions targeting resource-constrained environments, like the edge. Furthermore, we support EdgeNet, which is a cloud orchestration system for the edge cloud continuum.

Similarly, various network plugins can be utilized for each K8s distribution: (i) vanilla K8s: Flannel, Multus, Calico, WeaveNet, Cilium, Kube-Router, Kube-OVN, Antrea; (ii) K3s: Flannel, Calico, Cilium; (iii) K0s: Kube-Router, Calico; (iv) Microk8s: Calico, Flannel, Kube-OVN. In addition to the above CNI plugins, we have also implemented the support of L2S-M solution \cite{l2sm}, i.e., utilized by CODECO NetMA, within vanilla K8s (referred to as $k8s-l2sm-v1$). The considered network infrastructure (e.g., underlay or overlay) and the provided features (e.g., advanced security/encryption, programmability, network policies, etc.) vary among the plugins. Notably, Flannel, WeaveNet and Kube-Router provide simpler and more lightweight solutions; Calico, Cilium, Antrea and Kube-OVN support advanced and programmable features, while Multus allows for the simultaneous utilization of multiple plugins within a cluster.

Figs. \ref{plugins-metrics1} and \ref{plugins-metrics2} illustrate the resource consumption (CPU, RAM) and throughput performance of the supported network solutions across the different K8s distributions. The former focuses on the standard K8s flavor, while the latter evaluates the performance of networking solutions deployed in edge-related distributions, i.e., K3s, K0s, MicroK8s and EdgeNet. In this context, various communication conditions, i.e., “Idle”, “Pod-to-Pod” (p2p), and “Pod-to-Service” (p2s), are evaluated both for TCP and UDP transport layer protocols. The Kubernetes Network Benchmark (knb) tool\footnote{https://github.com/InfraBuilder/k8s-bench-suite} is utilized to obtain the aforementioned metrics in an intra-host deployment in the ATH testbed. The deployment of the distributions and respective plugins is carried out automatically by the framework resulting in identical clusters which comprise one master and two worker nodes (VMs). In \cite{iccn-plugins}, we document a relevant comprehensive performance evaluation. Consequently, our framework is capable of experimenting with multiple networking solutions over various edge-oriented K8s flavors.

\subsubsection{EdgeNet Deployment}
Here, we demonstrate the capability of CODECO Experimentation Framework to handle complex edge system deployments, such as the installation of EdgeNet, an open-source solution designed to extend K8s to the Edge\cite{edgenet1, edgenet3}. Such process involves installing EdgeNet prerequisites, e.g., creates kubeconfig and ssh-key secret certificates, configures wireguard, creates CRs for VPN peering, and installs the EdgeNet features implemented as CR Definitions and operators. Such deployment required also improvements in the EdgeNet software itself, e.g., to improve its autonomic deployment, compatibility and versioning. 

After the installation, a functional CODECO-EdgeNet K8s cluster is deployed, with supported functionalities, such as: i) \textit{Multi-tenancy}: role-based approach in the shared cluster, e.g., create and accept tenant and role requests, create slices, subnamespaces, allocate resource quotas, etc.; and 
ii) \textit{Multi-provider}: users from around the world can contribute nodes (VMs or physical ones) to the cluster, with a single command.

Fig \ref{fig:EdgeNet-output} illustrates the output of some basic $kubectl$ commands, showcasing the successful installation of EdgeNet software, the establishment of the cluster, and the node contribution processes. In particular, a modified version of the Definition File in \ref{exper-descr} is used for the installation of EdgeNet utilizing the specific EdgeNet-related automation features, followed by distributed nodes (VMs) 
that join this cluster with one-liner commands.
Subsequently, we complement Subsection \ref{plugins&distros} by evaluating the performance of the Antrea CNI plugin within this local EdgeNet cluster (referred to as $k8s-edgenet-antrea$) in Fig \ref{plugins-metrics1}.

\begin{figure*}
    \setlength{\columnsep}{0.2em} 
        \centering
        \fontsize{5.75pt}{10pt}
        \begin{verbatim}
athena@athm1:~$ kubectl get nodes -o wide
 NAME                    STATUS   ROLES            AGE     VERSION    INTERNAL-IP       OS-IMAGE             KERNEL-VERSION      CONTAINER-RUNTIME
 athm1                   Ready    control-plane,   28h     v1.23.17   83.212.134.25     Ubuntu 22.04.2 LTS   5.15.0-71-generic   containerd://1.6.24
                                  master
 athw2                   Ready    <none>           28h     v1.23.17   83.212.134.29     Ubuntu 22.04.2 LTS   5.15.0-71-generic   containerd://1.6.24
 athw3                   Ready    <none>           28h     v1.23.17   83.212.134.30     Ubuntu 22.04.2 LTS   5.15.0-71-generic   containerd://1.6.24
 gr-a-7230.edge-net.io   Ready    <none>           15m     v1.23.17   83.212.134.27     Ubuntu 22.04.2 LTS   5.15.0-86-generic   containerd://1.5.11
 gr-b-89c3.edge-net.io   Ready    <none>           108s    v1.23.17   195.251.209.229   Ubuntu 22.04.2 LTS   5.15.0-71-generic   containerd://1.5.11
 gr-b-abf4.edge-net.io   Ready    <none>           3m26s   v1.23.17   195.251.209.231   Ubuntu 22.04.2 LTS   5.15.0-71-generic   containerd://1.5.11

athena@athm1:~$ kubectl get nodecontributions
 NAME        ADDRESS           PORT   ENABLED   STATUS          AGE
 gr-a-7230   83.212.134.27     22     true      Node Accessed   16m
 gr-b-89c3   195.251.209.229   22     true      Node Accessed   2m34s
 gr-b-abf4   195.251.209.231   22     true      Node Accessed   4m13s
        \end{verbatim}
    \caption{EdgeNet installation and node contribution output commands.}
    \label{fig:EdgeNet-output}
\end{figure*}

\begin{figure*}[t]
\centering
   \begin{subfigure}[b]{0.32\textwidth}       \includegraphics[width=\textwidth]{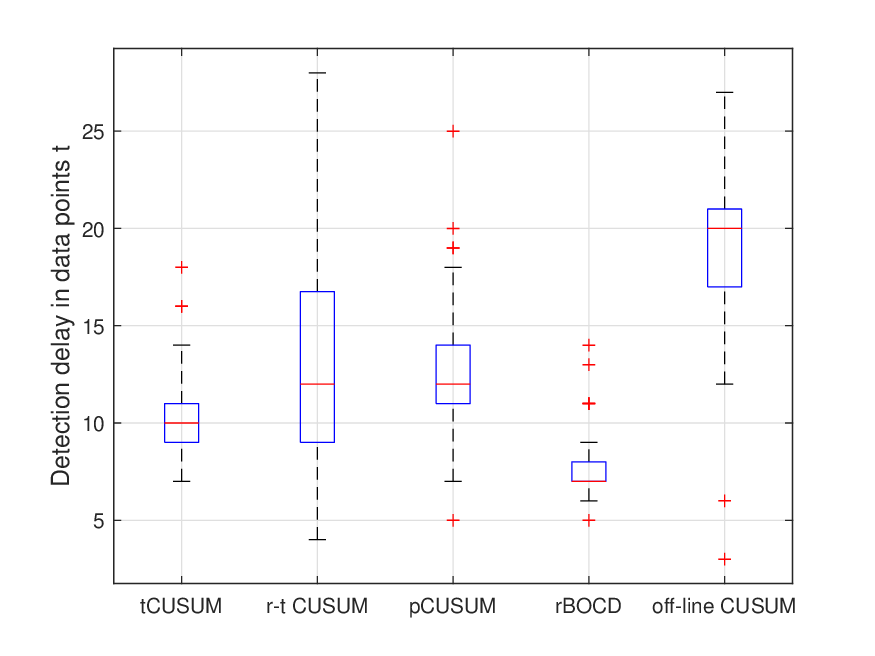}
   \caption{}
   \end{subfigure}%
   \begin{subfigure}[b]{0.32\textwidth}       \includegraphics[width=\textwidth]{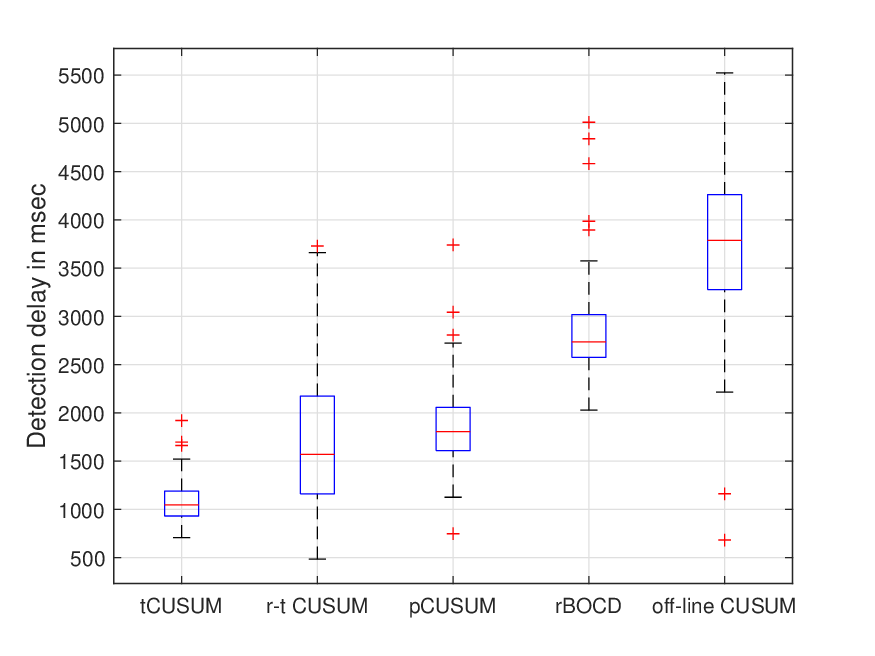}
   \caption{}
   \end{subfigure}%
   \begin{subfigure}[b]{0.32\textwidth}   \includegraphics[width=\textwidth]{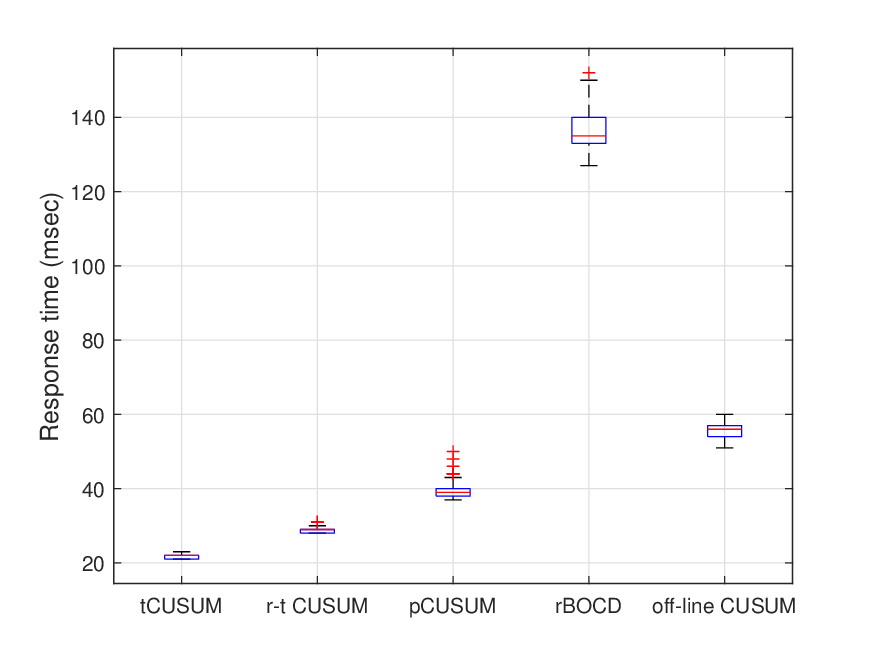}
   \caption{}
   \end{subfigure}
      \begin{subfigure}[b]{0.32\textwidth}   \includegraphics[width=\textwidth]{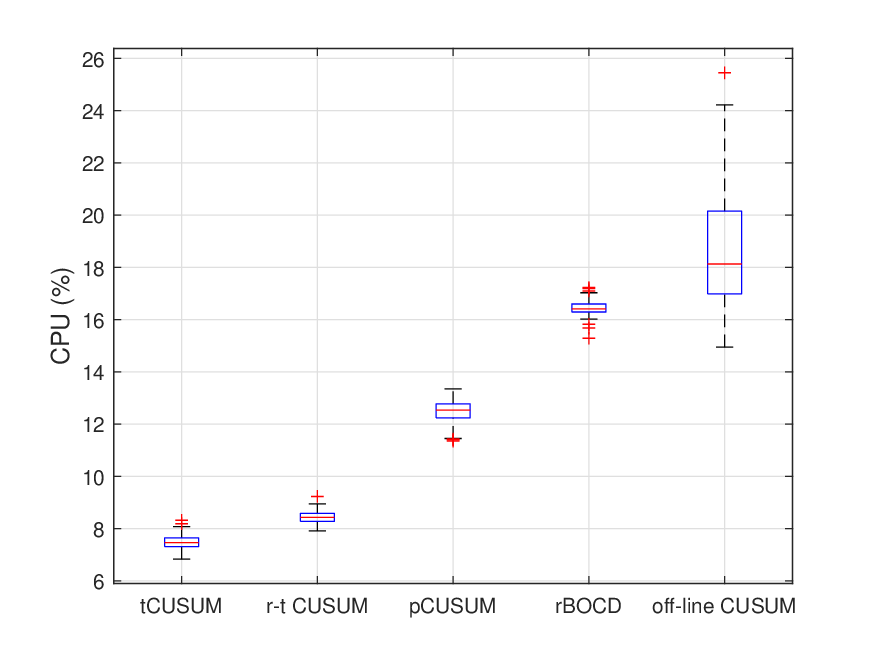}
   \caption{}
   \end{subfigure}
      \begin{subfigure}[b]{0.32\textwidth}   \includegraphics[width=\textwidth]{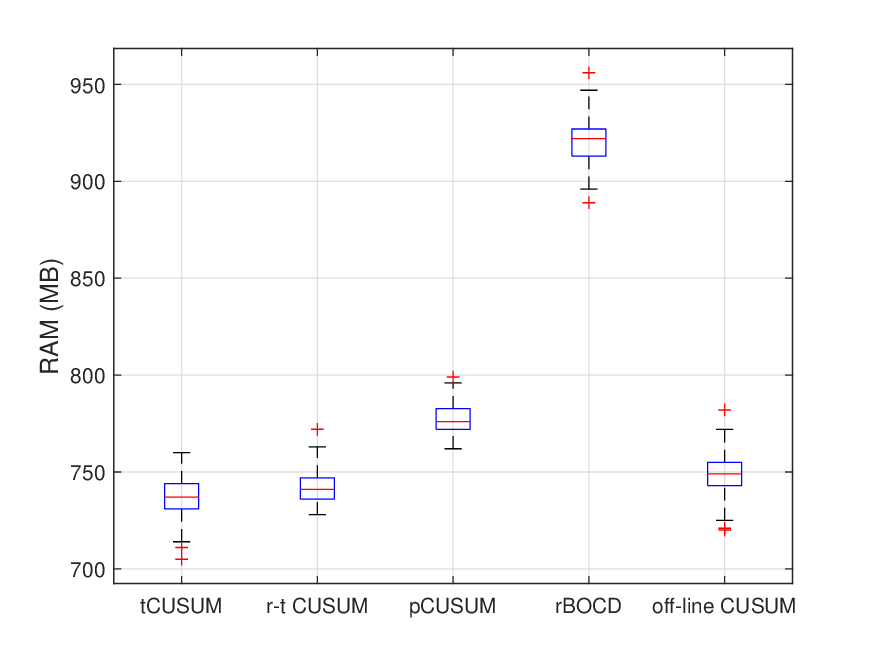}
   \caption{}
   \end{subfigure}
   \caption{Comparison of (a) detection gap in data points $t$, (b) detection gap in msec, (c) response time in msec, (d) CPU, and, (e) RAM consumption between the anomaly detection methods.}
   \label{anomaly-detection-fig}
\end{figure*}

\subsubsection{Anomaly Detection Workflows for the Edge}

Lastly, we demonstrate CODECO’s Experimentation Framework capabilities to assess the impact of ML algorithms in edge cloud environments, in terms of actual response time and resource utilization. Here, we focus on the evaluation of the following 5 AD detectors: i) typical CUSUM detector (tCUSUM) \cite{skaperas2019real}, ii) ratio-type CUSUM (r-t CUSUM) \cite{skaperas2019real}, iii) ARMA-based CUSUM (pCUSUM) \cite{aue2015reaction}, iv) restarted Bayesian (rBOCP) \cite{alami2020restarted}, and, iv) off-line CUSUM \cite{aue2013structural}. Note that the considered detectors operate online and are applied on the UOM testbed.
	
The AD methods are implemented as K8s argo workflows\footnote{https://github.com/argoproj/argo-workflows} (operated/removed on demand) and use a client-server application. In detail, the experimental parameters are defined through the workflows, e.g., the type of the AD method and the number of simultaneous clients to be deployed. The client-server communication is achieved through a REST API. The client transmits, in real-time, new data samples to the server, which employs the AD mechanisms and alerts the clients for anomalies. The servers also monitor the resource consumption of the AD mechanisms using an extension of knb tool. 

Fig. \ref{anomaly-detection-fig} illustrates the performance metrics for the considered AD approaches, assessed over 100 Monte Carlo simulations, for one client transmitting synthetic data following a piece-wise Gaussian distribution with one anomaly in the mean value. Particularly, Fig. \ref{anomaly-detection-fig}a depicts the common detection gap (in data points $t$) between the occurrence and the estimation of an anomaly. In addition, the CODECO Experimentation Framework extends the evaluation metrics to the actual detection gap and response time along with the CPU and memory consumption for each AD method, as shown in Figs. \ref{anomaly-detection-fig}b -- e. This provides valuable insights for practitioners about the performance of AD methods in real-world edge applications that cannot be revealed through typical statistical metrics. For instance, when comparing rBOCP with the online CUSUM procedures, the first concludes on a substantially higher actual detection gap in msec, despite providing the shortest detection gap in data points. A thorough evaluation of different AD methods in edge-cloud systems can be found in our previous work \cite{iccn}.

\section{Conclusions and Future work}\label{concl}
This paper introduced the open-source CODECO Experimentation Framework\footnote{We released the CODECO Experimentation Framework and our port of EdgeNet in https://gitlab.eclipse.org/eclipse-research-labs/codeco-project/\\
experimentation-framework-and-demonstrations} tailored for experimenting in the Edge. We presented its architecture, underscoring its capacity to facilitate the deployment of clusters, applications and experiments in a holistic and automated manner. Additionally, we demonstrated the framework's significance through three proof-of-concept examples.
Our future plans involve the support of additional pluggable K8s features and applications (e.g., the CODECO components), the investigation of multi-cluster solutions and its interconnection with external testbeds.

\section*{Acknowledgment}
This work is supported by the Horizon Europe CODECO Project under Grant number 101092696.

\bibliography{codeco-experiments.bib}
\bibliographystyle{IEEEtran}

\end{document}